\documentstyle[12pt]{article}                                             

       \def\de{depth}

\let\a=\alpha \let\be=\beta \let\g=\gamma \let\de=\delta                 
\let\e=\varepsilon                  
  \let\la=\lambda \let\m=\mu               
\let\n=\nu   \let\r=\rho \let\s=\sigma                
 \let\t=\perp                                               
\let\ph=\varphi  \let\PH=\Phi                     
                              
\let\La=\Lambda  \let\D=\Delta                               
                                              
\def\0{\over } \def\1{\vec }     \def\2{{1\over2}} \def\4{{1\over4}}     
\def\5{\bar }  \def\6{\partial } \def\7#1{{#1}\llap{/}}                  
\def\8#1{{\textstyle{#1}}}       \def\9#1{{\bf {#1}}}                    
 \def\llp{\hbox to 0pt{\hss /\hskip1.5pt}} 
\def\llo{\hbox to 0.2pt{\hss /}} \def\llq{\hbox to 0pt{\hss /\hskip0.5pt}}   
\def\so{\supset\hbox to 0pt{\hss $\displaystyle -$\hskip1pt}}

\def\<{\langle } \def\>{\rangle }

\let\nn=\nonumber                                                         
\def\bea{\begin{eqnarray}} \def\eea{\end{eqnarray}}                       
\def\beann{\begin{eqnarray*}} \def\eeann{\end{eqnarray*}}                
\def\beq{\begin{equation}} \def\eeq{\end{equation}}

\newcommand{\Dl}{D\!\!\!\!\!\raisebox{1.5ex}{$\leftarrow$}}

\newcommand{\Gs}{G\!\!\!\!/}
\newcommand{\Ds}{D\!\!\!\!/}
\newcommand{\ls}{l\!\!/}
\newcommand{\ks}{k\!\!\!/}
\newcommand{\bls}{\bar{l}\!\!/}
\newcommand{\bps}{\bar{p}\!\!/}
\newcommand{\psu}{p\!\!/}
\newcommand{\pts}{\tilde{p}\!\!/}
\newcommand{\lts}{\tilde{l}\!\!/}
\newcommand{\qs}{q\!\!\!/}
\newcommand{\epsilons}{\epsilon\!\!/}
\newcommand{\bp}{\bar{p}}
\newcommand{\bl}{\bar{l}}

\newcommand{\pdl}{\partial\!\!\!\!\raisebox{1.5ex}{$\leftarrow$}}

\date{}
\title{
{\large\rm DESY 96-126}\hfill{\large\tt ISSN 0418-9833}\\
{\large\rm SLAC-PUB-7204}\hfill\vspace*{0cm}\\
\hfill\vspace*{0cm}\\
\hfill\vspace*{2.5cm}\\
Gluon Radiation in Diffractive Electroproduction}
\author{W. Buchm\"uller, M. F. McDermott\\
\vspace{3.0\baselineskip}                                               
{\normalsize\it Deutsches Elektronen-Synchrotron DESY, 22603 Hamburg,
  Germany}
\\[.2cm]
and\\[.2cm]
A. Hebecker\\
{\normalsize\it Stanford Linear Accelerator Center, P.O.~Box 4349, MS 81,
CA 94309, USA}
\vspace*{2cm}\\                     
}                                                                          
\input psfig 
\begin{document}                                                  

\setlength{\baselineskip}{18pt}                                     
\maketitle  
\begin{abstract}
\noindent
Order $\a_s$-corrections to the diffractive structure functions $F_L^D$
and $F_2^D$ at large $Q^2$ and small $x$ are evaluated in the semiclassical
approach, where the initial proton is treated as a
classical colour field. The diffractive final state contains a fast gluon
in addition to a quark-antiquark pair. Two of these partons may have large
transverse momentum. Our calculations lead to an intuitive picture of
deep-inelastic diffractive processes which is very similar to Bjorken's 
aligned-jet model. Both diffractive structure functions contain leading 
twist contributions from high-$p_{\perp}$ jets.

\end{abstract} 
\thispagestyle{empty}
\newpage

\section{Introduction}

The large-rapidity-gap events, observed in deep inelastic scattering at 
HERA \cite{zeus,h1}, are a puzzling phenomenon. As the data show, the cross 
section for these `diffractive' events is not suppressed at large values 
of $Q^2$ relative to the inclusive cross section, i.e., these events 
represent a `leading twist' contribution to the inclusive structure function. 
However, since the initial proton escapes essentially unscathed, they cannot
be described by ordinary perturbation theory within the parton model. 
One may therefore hope that the study of deep inelastic diffractive processes 
will lead to a deeper understanding of the interplay between `soft' and 
`hard' processes in QCD.

A theory of hard diffractive processes, which does not yet exist, has 
to provide a clean separation between the perturbative and the 
non-perturbative aspects of the scattering process. In particular, 
it should explain why the large-rapidity-gap events are a `leading twist' 
phenomenon, and it should yield a quantitative description of at least some 
aspects of the diffractive structure functions. At present the basic mechanism
responsible for hard diffraction has not yet been unequivocally identified. 
Several phenomenological approaches have been developed which may be 
partially interrelated. These include the aligned-jet model \cite{bj1} 
(whose origin dates back before the development of QCD), the idea of a 
`pomeron' structure function \cite{ingel}, perturbative 2-gluon
exchange \cite{ryskin} and partonic models supplemented by soft 
colour interactions \cite{bh1,rath}.  

Recently, a semiclassical approach to diffractive deep inelastic
scattering has been proposed \cite{bh2}. In the proton rest frame
the dominant process is the dissociation of the
virtual photon into a quark antiquark pair, and further gluons and
quark pairs generated by radiation, which then interact with the 
proton\footnote{For a discussion and references, see \cite{dok}.}.
The basic idea of the semiclassical approach is to treat the proton at 
small values of $x$, where the momentum transfer is generally small, 
as a classical colour field localized within a radius 
$1/\Lambda_{QCD}$. Deep inelastic scattering is then viewed as
scattering of this system of fast quarks and gluons 
off the classical colour field. If this system happens to be in a colour
singlet state after the interaction one obtains a `diffractive' final
state with a large rapidity gap between the fast proton and the diffractive
hadronic system. Otherwise, an ordinary non-diffractive final state is
produced.

In \cite{bh2} the dominant contributions to inclusive and diffractive
structure functions, due to a virtual quark-antiquark initial state,
were calculated. The results depend on non-perturbative averages over
the colour field inside the proton. Here we
compute ${\cal O}(\alpha_s)$-corrections to these results by taking the
radiation of a gluon in the initial state into account. This allows
us to discuss the cross section for the production of high-$p_{\perp}$
jets. Furthermore, we will clarify the connection of our approach to
previous work on pair production in external fields based on light-cone
quantization \cite{soper}.

The paper is organized as follows. In Sect.~2 we discuss the applicability
of the semiclassical approach to high energy diffractive scattering.
In Sect.~3 we then rederive the results obtained in \cite{bh2} in a way
closely related to the techniques used in \cite{soper}. This calculation 
is extended to the case of an initial quark-antiquark-gluon system in
Sect.~4 with emphasis on the production of high-$p_{\perp}$ jets. Our
results are summarized in Sect.~5, and some details of the calculations 
we performed are explained in the appendices.
  
\section{Semiclassical approach to high-energy scattering}\label{sc}

The subject of the present paper is the analysis of diffractive deep 
inelastic scattering in terms of the production of quark-antiquark and 
quark-antiquark-gluon final states. These processes are considered in the 
small-$x$ or high-energy limit, where a considerable part of the total 
$\gamma^*p$ cross section is diffractive. The gluon densities are known 
to grow rapidly in this small-$x$ region. Therefore, as already discussed 
in \cite{bh2}, a description of the proton in terms of a classical colour 
field should be adequate. This is closely related to the description of 
high energy processes in terms of Wilson lines, which has been discussed 
by several authors \cite{nacht,kl}. In this section the basis of 
the semiclassical approach shall be discussed in more detail. It is similar 
to the method developed by Balitsky \cite{bal} for small-$x$ deep inelastic 
scattering. However, in the following more emphasis shall be placed on the 
connection with the proton wave functional in the Schr\"odinger picture. 

To keep the discussion as simple as possible, consider first the elastic 
scattering of a quark off a proton. Although this process is unphysical 
since quarks are confined, it can serve to illustrate the method of  
calculation. Therefore, in the following confinement is ignored and 
quarks are treated as asymptotic states. The generalization to the 
physical case of diffractive electroproduction is straightforward and will 
be discussed subsequently. 

\begin{figure}[h]
\begin{center}
\parbox[b]{8cm}{\psfig{width=7cm,file=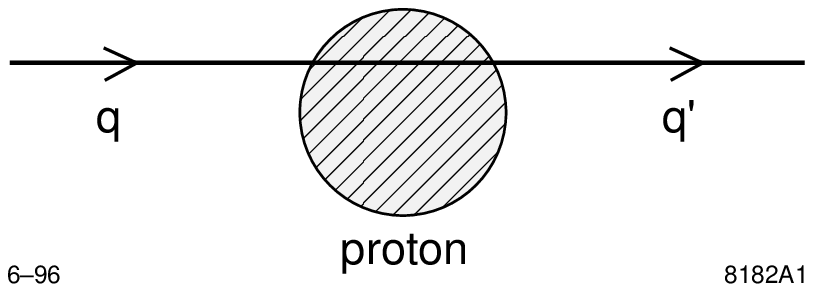}}\\
\end{center}
\refstepcounter{figure}
\label{qp}
{\bf Fig.\ref{qp}} Scattering of a point-like quark off the proton bound 
state. 
\end{figure}

The scattering of a point-like quark with initial momentum $q$ off a 
relativistic bound state with initial momentum $p$ is illustrated in 
Fig.~\ref{qp}. Let $m_p$ be the proton mass, and $s$ and $t$ the usual 
Mandelstam variables for a $2\to 2$ process. In the high-energy limit, 
\hspace{.2cm}$s\gg t,\, m_p^2$,\hspace{.2cm} the 
contribution from the annihilation of the incoming quark with a constituent 
quark of the proton is negligible. The amplitude is dominated by diagrams 
with a fermion line going directly from the initial to the final quark 
state. Therefore, the proton can be described by a Schr\"odinger wave 
functional $\Phi_p[A]$ (cf.~\cite{luscher}) depending on the gluon field only.
Quarks are integrated out, yielding a modification of the gluonic action. 

For a scattering process the amplitude can be written in the proton rest 
frame as
\beq \label{qamp}
<\!q'p'|qp\!>=\lim_{T\to\infty}\int DA_T DA_{-T} \Phi_{p'}^*[A_T] 
\Phi_p[A_{-T}]\int_{A_{-T}}^{A_T}DA\, e^{iS[A]}<\!q'|q\!>_A\, . 
\eeq 
Here the fields $A_{-T}$ and $A_T$ are defined on three-dimensional surfaces 
at constant times $-T$ and $T$, and $A$ is defined in the four-dimensional 
region bounded by these surfaces. The field $A$ has to coincide with 
$A_{-T}$ and $A_T$ at the boundaries and the action $S$ is defined by an 
integration over the domain of $A$. The amplitude $<\!q'|q\!>_A$ 
describes the scattering of a quark by the given external field $A$.

The initial state proton, having well defined momentum $\vec{p}$, 
is not well localized in space. However, the dominant field configurations
in the proton wave functional are localized on a scale $\La \sim \La_{QCD}$.
Also the field configurations $A(\vec{x},t)$, which interpolate between
initial and final proton state, are localized in space at each time $t$.
Assume that the incoming quark wave packet is localized such that it passes
the origin $\vec{x}=0$ at time $t=0$. At this instant the field configuration
$A(\vec{x},t)$ is centered at
\beq 
\vec{x}[A]\equiv\int d^3\vec{x} E_A(\vec{x})\cdot\vec{x}\Bigg/ 
\int d^3\vec{x} E_A(\vec{x})\, , 
\eeq 
where $E_A(\vec{x})$ is the energy density of the field $A(\vec{x},t)$ at 
$t=0$. The amplitude (\ref{qamp}) can now be written as
\beq 
<\!q'p'|qp\!>=\lim_{T\to\infty}\int d^3\vec{x}\int DA_T DA_{-T} 
\Phi_{p'}^*\Phi_p\int_{A_{-T}}^{A_T}DA\, e^{iS}
\delta^3(\vec{x}[A]-\vec{x})<\!q'|q\!>_A\, . 
\eeq 
Using the transformation properties under translations,
\beq
<\!q'|q\!>_{L_{\vec{x}}A}=e^{i(\vec{q}-\vec{q}\,')\vec{x}}<\!q'|q\!>_A
\quad,\quad\Phi_{p'}^*[L_{\vec{x}}A]\Phi_p[L_{\vec{x}}A]=
e^{i(\vec{p}-\vec{p}\,')\vec{x}}\Phi_{p'}^*[A]\Phi_p[A]\,\, ,
\eeq
where 
\beq
L_{\vec{x}}A(\vec{y})\equiv A(\vec{y}-\vec{x})\, ,
\eeq
one obtains, 
\beq
<\!q'p'|qp\!>=2m_p(2\pi)^3\delta^3(\vec{p}\,'+\vec{q}\,'-\vec{p}-\vec{q})
\, \int_A <\!q'|q\!>_A\, .
\eeq
Here $\int_A$ denotes the operation of averaging over all field 
configurations contributing to the proton state which are localized at 
$\vec{x}=0$ at time $t=0$. It is defined by 
\beq
\int_A F\equiv\frac{1}{2m_p}\lim_{T\to\infty}
\int DA_T DA_{-T} \Phi_{p'}^*\Phi_p\int_{A_{-T}}^{A_T}DA\, e^{iS}
\delta^3(\vec{x}[A])F[A]
\eeq
for any functional $F$. The normalization $\int_A\ 1=1$ follows from 
\beq
<\!p'|p\!>=2p_0(2\pi)^3\delta^3(\vec{p}\,'-\vec{p}\,)\ .
\eeq

More complicated processes can be treated in complete analogy as long as the
proton scatters elastically. In particular, the above arguments apply to 
the creation of colour singlet quark antiquark pairs \cite{bh2} 
\beq
<\!q\bar{q}p'|\gamma^*p\!>=2m_p(2\pi)^3\delta^3(\vec{k}_f-\vec{k}_i)\ 
\int_A <\!q\bar{q}|\gamma^*\!>_A\, ,\label{qqa}
\eeq
where $\vec{k}_i$ and $\vec{k}_f$ are the sums of the momenta in the initial 
and final states respectively.  
The generalization of this simplest diffractive process to 
a process with an additional fast final state gluon, $\gamma^*\to q\bar{q}
g$, will be given in Sect.~\ref{qqg}. In contrast to the quark-proton 
scattering discussed above, here a colour neutral state is scattered off the 
proton. Therefore no immediate contradiction with colour confinement arises. 
However, it has to be assumed that the hadronization of the produced 
partonic state takes place after the interaction with the proton, 
which is described in terms of fast moving partons.

When calculating the cross section from Eq.~(\ref{qqa}) the square of the 
momentum conserving $\delta$-function translates into one momentum 
conserving $\delta$-function using Fermi's trick. This $\delta$-function
disappears after the momentum integration for the final state proton, 
resulting in a cross section formula identical to scattering by an external 
field. 

{}From the above discussion a simple recipe for the calculation of 
diffractive processes at high energy follows: 

The partonic process is calculated in a given external colour field, 
localized at $\vec{x}=0$ at time $t=0$. The weighted average over all 
colour fields contributing to the proton state is taken on the amplitude 
level. We assume that the typical contributing field is smooth on a 
scale $\Lambda$ and is localized in space on the same scale $\Lambda$. 
Finally, the cross section is calculated using standard formulae for 
the scattering off an external field. 

For the comparison of diffractive and inclusive structure functions it is 
important to know whether the above semiclassical picture  also applies 
to non-singlet partonic final states, e.g. to the production of a 
colour-octet $q\bar{q}$-pair. In this situation the definition of some 
analog of the `averaging'-operator in terms of a path integral and the 
proton wave functional is not obvious. The problem is that the fragmentation 
will in general involve both the diffractively  produced final state and the 
proton remnant. Below we shall assume that such processes can still 
be described by calculating the partonic amplitude in a smooth localized 
colour field and by taking an appropriate average over different field 
configurations at the end. 

\section{Quark pair production in a colour field}\label{qq}

\begin{figure}[h]
\begin{center}
\parbox[b]{10cm}{\psfig{width=10cm,file=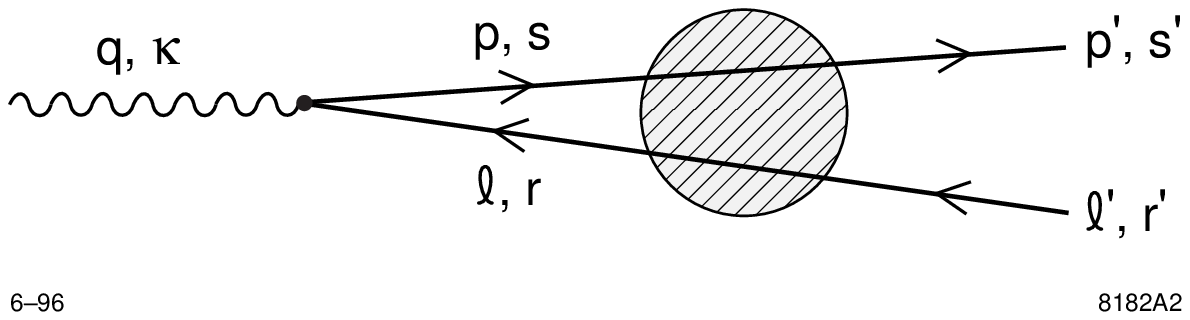}}\\
\end{center}
\refstepcounter{figure}
\label{dqq}
{\bf Fig.\ref{dqq}} Quark pair production in the colour field of the proton.
\end{figure}

The basic process in deep inelastic scattering, viewed in the proton
rest frame, is the dissociation of a virtual photon into a
quark antiquark pair which then interacts with the 
proton. The corresponding scattering amplitude reads (cf.~\cite{bh2})
\beq\label{amp}
S_{\mu} = i e \int d^4x e^{-iqx} \bar{\psi}_u(x)\gamma_{\mu}
\psi_v(x)\ ,
\eeq
where $\psi_u$ and $\psi_v$ are the wave functions of the outgoing
quark and antiquark depending on the momenta $p'$ and $l'$,
respectively, and $e$ is the electric charge (cf.~Fig.~\ref{dqq}). 

In the semiclassical approximation the interaction with the proton is 
treated as scattering in a classical colour field $G_{\mu}(x) = T^a 
G_{\mu}^a(x),\,T^a = {1\over 2} \lambda^a$. The quark and antiquark wave 
functions are solutions of the Dirac equation with the colour field. Hence, 
they satisfy the integral equations
\bea\label{int}
\psi_v(x) &=& \psi^{(0)}_v (x)   - \int d^4x' S_F(x-x') \Gs(x') 
\psi_v(x')\ ,\\ \bar{\psi}_u(x) &=& \bar{\psi}^{(0)}_u (x) -\int d^4x' 
\bar{\psi}_u(x') \Gs(x') S_F(x'-x)\ .
\eea 

Here $\psi^{(0)}_v$  and $\bar{\psi}^{(0)}_u$ are solutions of the
free Dirac equation. In the following we will explicitly consider
only those contributions to $S_{\mu}$ in which both the quark and
antiquark interact with the field $G_{\mu}$.  

Inserting these equations into the amplitude (\ref{amp}), using the
Fourier decomposition of the free propagator $S_F$ and performing the
integration over the position of the photon vertex, one obtains
\beq\label{amp2}
S_{\mu} = -i e \int {d^4p\over (2\pi)^4} \int d^4x
\bar{\psi}_u(x)\Gs(x) e^{-ipx} {1\over \psu - m} \gamma_{\mu}
{1\over \ls + m} \int d^4y e^{-ily} \Gs(y) \psi_v(y)\ ,
\eeq
where $l=q-p$. Explicit expressions for $\bar{\psi}_u$
and $\psi_v$ have been obtained in a high-energy expansion for 
an arbitrary soft colour field \cite{bh2}.

We are interested in deep inelastic scattering at small $x$, where
quark and antiquark have large momenta in the proton rest frame.
Hence, the propagators in Eq.~(\ref{amp2}) can be treated in a
high-energy approximation. It is convenient to introduce light-cone
variables, e.g.,
\beq
l_+=l^0+l^3\ ,\ l_-=l^0-l^3\ ,\ \bl_-={l_{\perp}^2+m^2\over l_+}\ .
\eeq
Here $\bl_{\mu}$ denotes the momentum vector whose ``--''-component
satisfies the mass shell condition. The propagators in Eq.~(\ref{amp2})
can be written as
\bea
{1\over \ls +m} &=& \frac{\sum_r v_r(\bl)\bar{v}_r(\bl)}
{l_+(l_--\bl_-) + i\epsilon} + \frac{\gamma_+}{2 l_+}\ ,\label{prop1}\\
{1\over \psu - m} &=& \frac{\sum_s u_s(\bp) \bar{u}_s(\bp)}
{p_+(p_--\bp_-) + i\epsilon} + \frac{\gamma_+}{2 p_+}\ .\label{prop2}
\eea
To obtain the first term in a high energy expansion
of the scattering amplitude $S_{\mu}$ one can drop the terms
proportional to $\gamma_+$ in Eqs.~(\ref{prop1}),(\ref{prop2}). 

In the high energy expansion the leading term for the wave functions 
$\psi_v$ and $\bar{\psi}_u$ is the product of a non-abelian eikonal factor 
and a plane wave solution of the Dirac equation \cite{nacht}. 
The scattering amplitude (\ref{amp2}) then takes the 
form of a product of the photon-quark-antiquark vertex, propagator factors
and two matrix elements of an effective gluon vertex between on-shell
spinors. These matrix elements describe the elastic scattering between
the high energy (anti)quark and the proton, as discussed in Sect.~2. 
They are evaluated in Appendix A.
For a soft gluon field inside the proton one finds for the matrix element
of the antiquark
\bea
T_{r,r'}(l,l') &=& \int d^4y\ \bar{v}_r(\bar{l}) e^{-ily}\Gs(y) \psi_v(y)\nn\\
&\simeq& -2\pi i\ 2 l_+ \delta_{rr'} \delta(l_+-l_+') 
         \left(\tilde{F}(l_{\perp}-l'_{\perp}) - 
               (2\pi)^2 \delta^2(l_{\perp}-l'_{\perp})\right)\ ,
\eea
where 
\bea
\tilde{F}(l_{\perp}-l'_{\perp}) &=&
\int_{y_{\perp}}\ e^{i(l_{\perp}-l'_{\perp})y_{\perp}} F(y_{\perp})\ , \nn\\
F(y_{\perp}) &=& P \exp{\left({i\over 2} \int_{-\infty}^{\infty} dy_+ 
G_-(y_+,y_-,y_{\perp})\right)}\ .
\eea
$F(y_{\perp})$ is the eikonal factor of the antiquark trajectory. 
If the colour field of the proton is `soft'
the dependence on $y_-$ can be neglected. For the corresponding matrix
element of the quark one finds
\bea
\int d^4y\ \bar{\psi}_u(y) \Gs(y) u_s(\bp) e^{-ipy}
&\!\!\simeq\!\!& 2\pi i\ 2 p_+ \delta_{s's} \delta(p_+'-p_+) 
\left(\tilde{F}^{\dagger}(p'_{\perp} - p_{\perp}) - 
      (2\pi)^2 \delta^2(p'_{\perp}-p_{\perp})\right)\nn\\
&\!\!=\!\!& T^{\dagger}_{s's} (p',p)\ .
\eea

Inserting these matrix elements into Eq.~(\ref{amp2}) and adding the
contributions where one of the particles is not scattered one obtains,
after performing the $p_-$-integration,
\bea\label{amp3}
\bar{S}_{\mu} &=&  {e\over \pi}\ q_+ \delta(q_+ - p_+' - l_+')
  \int d^2l_{\perp}\ {\alpha(1-\alpha)\over N^2 + l_{\perp}^2}
  \ \bar{u}_{s'}(\bar{p})\gamma_{\mu}v_{r'}(\bar{l})\nn\\
&&\qquad\qquad\qquad \left(\tilde{F}^{\dagger}(p'_{\perp}-p_{\perp})
   \tilde{F}(l_{\perp}-l'_{\perp}) - (2\pi)^4 \delta^2(p'_{\perp}-p_{\perp})
   \delta^2(l_{\perp}-l'_{\perp})\right)\nn\\
&\equiv& - 4\pi\ \delta(q_+ - p'_+ - l'_+)\ T_{\mu}\ ,
\eea
where
\beq
p'_+ = (1-\alpha)\ q_+\ ,\ l'_+ = \alpha\ q_+\ ,\ 
N^2 = \alpha (1 - \alpha) Q^2 + m^2\ .
\eeq
In the following it will turn out to be useful to restrict the integrand
in Eq.~(\ref{amp3}) to configurations with $\alpha < 1/2$,
where the quark is faster than the antiquark.

The non-perturbative interaction with the proton is contained in $\tilde{F}$,
the Fourier transform of the eikonal factor. The product of eikonal factors 
appearing in Eq.~(\ref{amp3}) may be expressed as Fourier transform with 
respect to the transverse distance between quark and antiquark. Introducing
\beq\label{wdef}
W_{x_{\perp}}(y_{\perp}) = F^{\dagger}(x_{\perp})F(x_{\perp}+y_{\perp}) - 1\ ,
\eeq
one has
\bea
\tilde{F}^{\dagger}(p'_{\perp}-p_{\perp}) \tilde{F}(l_{\perp}-l'_{\perp})
&=& \int_{x_{\perp},z_{\perp}}\ e^{i(p_{\perp}-p'_{\perp})x_{\perp}}
    e^{i(l_{\perp} - l'_{\perp}) z_{\perp}}
    F^{\dagger}(x_{\perp}) F(z_{\perp})\nn\\
&=& \int_{x_{\perp},y_{\perp}}\ e^{-i\Delta_{\perp}x_{\perp}}
        e^{i(l_{\perp}-l'_{\perp})y_{\perp}}\
        \left(W_{x_{\perp}}(y_{\perp})+1\right)\nn\\
&=& \int_{x_{\perp}}\ e^{-i \Delta_{\perp}x_{\perp}}\
   \left(\tilde{W}_{x_{\perp}}(l_{\perp}-l'_{\perp}) + 
         (2\pi)^2 \delta^2(l_{\perp}-l'_{\perp})\right)\ .
\eea
Here $\Delta_{\perp}=p'_{\perp}+l'_{\perp}$ is the transverse momentum
transfer from the proton. We assume that the colour field of the proton is 
smooth on a scale $\Lambda$, which implies that $\tilde{W}_{x_{\perp}}
(k_{\perp})$ falls off exponentially with increasing 
$k_{\perp}^2/\Lambda^2$. 

Note, that our treatment of the background field assumes the
factorization of soft gluonic physics associated with the proton state 
and higher order $\alpha_S$-corrections associated with the photon
wave function. This property is not proved in the present paper.
Nevertheless, the calculations of the following section illustrate for
the specific case of high-$p_\perp$ jets in diffraction, how such 
$\alpha_S$-corrections can be implemented in the present approach.

Cross sections and structure functions can now be evaluated in the standard
manner (cf.~\cite{bh2}). The deep inelastic cross sections are given by
\beq\label{dsig}
d \sigma_{\mu\nu} = {2\pi\over q_+}\ T^*_{\mu}T_{\nu}\ d\PH^{(2)}\ ,
\eeq
where $d\PH^{(2)}$ is the phase space factor. Different projections with 
respect to Lorentz and colour structure yield longitudinal and transverse, 
diffractive and inclusive structure functions.

Consider first the inclusive longitudinal structure function $F_L$. We use 
the conventional kinematic variables $\xi=x/\beta=x(Q^2+M^2)/Q^2$,
where $M^2$ is the invariant mass of the produced quark-antiquark pair and 
$m=0$. A straightforward calculation, described in Appendix B, 
yields 
\bea
dF_L &=& {Q^2\over \pi e^2}\ d\sigma_L \nn\\
     &=& {4\ Q^6\over (2\pi)^7 \beta}\ {d\xi\over \xi}\ d\alpha 
         (\alpha(1-\alpha))^3\
         \int_{x_{\perp}} \left|\int d^2 l_{\perp} 
         \frac{\tilde{W}_{x_{\perp}}(l_{\perp}-l'_{\perp})}{N^2 + l_{\perp}^2}
         \right|^2\ .
\eea
Note that only a single integration over transverse coordinates occurs.
This is a consequence of the $\delta$-function induced by the phase space
integration over $\Delta_{\perp}$ in Eq.~(\ref{dsig}). Since the form factor
$\tilde{W}_{y_{\perp}}(l_{\perp}-l'_{\perp})$ falls off exponentially with
increasing momentum transfer, one can expand the integrand around 
$l_{\perp}=l'_{\perp}$. This leads to the final result
\bea\label{fl}
F_L &=& {2\over \pi^3}\ \int_x^1{d\xi\over \xi}\ 
                        \int_0^{1/2} d\alpha\ \beta^2(1-\beta)
         \int_{x_{\perp}} \mid\partial_{\perp}W_{x_{\perp}}(0)\mid^2 \nn\\
     &=& {1\over 6\pi^3}\ \int_{x_{\perp}} 
          \mid\partial_{\perp}W_{x_{\perp}}(0)\mid^2\ .
\eea

The inclusive transverse structure function $F_T$ can be evaluated in a
similar way. In the perturbative region $\alpha > \Lambda^2/Q^2$, where
the antiquark is sufficiently fast, one obtains
\bea\label{ft0}
dF_T &=& {Q^2\over \pi e^2}\ d\sigma_T  = dF_2 - dF_L \nn\\
     &=& {1\over 8\pi^3}\ {d\xi\over \xi}\ d\alpha\ 
         \frac{\alpha^2+(1 - \alpha)^2}{\alpha(1-\alpha)}\ 
         \beta(\beta^2+(1-\beta)^2)\ 
         \int_{x_{\perp}} \mid\partial_{\perp}W_{x_{\perp}}(0)\mid^2\ .
\eea
The integration over $\alpha$ above the infrared cutoff 
$\alpha_{min}=\Lambda^2/Q^2$ yields
\beq\label{infra}
\int_{\alpha_{min}}^{1/2}\ d\alpha\ 
\frac{\alpha^2+(1 - \alpha)^2}{\alpha(1-\alpha)}
\simeq \ln{{Q^2\over \Lambda^2}} - 1\ .
\eeq

{}From the general expression for $F_T$ (cf. Appendix B) one can easily
see that there is also a non-perturbative contribution from the range
$\alpha \leq \Lambda^2/Q^2$ which, to leading order in $1/Q^2$, is given
by a function of $\beta$ only. We thus obtain the final result
\bea\label{ft}
F_T &=& {1\over 4\pi^3}\ \int_x^1{d\xi\over \xi}\ 
         \left( \beta(\beta^2+(1-\beta)^2)\  
         \left(\ln{{Q^2\over \Lambda^2}} - 1\right)\ 
         \int_{x_{\perp}} \mid\partial_{\perp}W_{x_\t}(0)\mid^2\
         + f(\beta)\right)\nn\\
    &=& {1\over 6\pi^3}\ \ln{{Q^2\over \Lambda^2}}\   
         \int_{x_{\perp}} \mid\partial_{\perp}W_{x_\t}(0)\mid^2\ +\ C\ ,
\eea
where $C$ is an unknown constant. The lower region of the 
$\alpha$-integration, responsible for the constant $C$, is dominated by 
$\alpha \sim \Lambda^2/Q^2 $ (see Eq.~(\ref{dft}) of Appendix B). 
In the small-$x$ limit this means that in the dominant configurations
the `soft' antiquark will still be sufficiently fast for the eikonal 
approximation to apply. Results for the production of 
electron-positron pairs in an electromagnetic field, which are completely
analogous to Eqs.~(\ref{fl}) and (\ref{ft}), have previously been
obtained by Bjorken, Kogut and Soper \cite{soper} using light-cone
quantization.

Let us finally evaluate the diffractive structure functions. In our approach
they are determined by the projection onto a colour singlet final state
which corresponds to the substitution
\beq
W_{x_{\perp}}(y_{\perp}) \rightarrow {1\over \sqrt{3}}\ 
          \mbox{tr}[W_{x_{\perp}}(y_{\perp})]\ .
\eeq
There is no non-perturbative contribution to $F_L$ to leading order
in $1/Q^2$. Further, from the definition of the colour matrix $W_{x_\t}(y_\t)$
it is clear that
\beq
\partial_{\perp}\mbox{tr}[W_{y_{\perp}}(0)] = 0\ .
\eeq
This immediately implies
\beq
F_L^D(x,Q^2,\xi) = 0\ .
\eeq
Like the transverse inclusive structure function, the transverse diffractive
structure function also has a non-perturbative contribution. The result can 
be written in the form
(cf. Appendix B)
\beq\label{f2d1}
F_2^D(x,Q^2,\xi) = {\beta\over \xi} \bar{F}(\beta)\ ,
\eeq
with
\beq\label{f2d2}
\bar{F}(\beta) = {4\over 3(2\pi)^7}\int d\rho \rho^3\ \int_{x_{\perp}}
\left|\int d^2k_{\perp}\ \frac{k_{\perp} + e_{\perp}\rho\sqrt{1-\beta}}
{\beta\rho^2 + (k_{\perp}+e_{\perp}\rho\sqrt{1-\beta})^2}
\mbox{tr}[\tilde{W}_{x_{\perp}}(k_{\perp})]\right|^2\ .
\eeq
Here $e_\t$ is an arbitrary unit vector whose direction is irrelevant
due to rotational invariance. The function $\bar{F}(\beta)$ approaches a 
finite limit as $\beta \rightarrow 0$, as well as $\beta \rightarrow 1$.

The results of this section essentially coincide with \cite{bh2}. There
is, however, a difference concerning the $\beta$-spectrum of the
diffractive contribution. In the space-time picture of \cite{bh2} the
change of direction of the outgoing quarks by their interaction with
the proton was not treated sufficiently accurately. This oversimplification
led to the conclusion that the $\beta$-spectrum depends on the proton
structure only via two unknown constants. The present calculation shows
that the $\beta$-spectrum is entirely non-perturbative and can only be 
given in terms of an integral which depends on the proton field.

The results for $F_2$ and $F_L$ correspond to the
perturbative contribution of photon-gluon fusion with a gluon density
$x g(x) =$ const., i.e. a classical bremsstrahl spectrum. The
diffractive structure function $F_2^D$ is due to purely non-perturbative
contributions, where either the quark or the antiquark are soft. 
This is analogous to modern views of the aligned-jet model 
\cite{fs,frank,bj2,elro}. 
We also note that the slope of $F_2$ in $x$ 
is larger by one unit than the slope 
of $F_2^D$ in $\xi$. This property of the structure functions has 
previously been discussed in \cite{bh1,bh2,b0}. 

\section{Radiation of an additional gluon}\label{qqg}

\subsection{Amplitude}\label{ampl}

In this section the process $\gamma^*\to q\bar{q}g$ in an external colour 
field is calculated in the kinematical region with two final state partons 
having high transverse momentum. This extends the analysis of the previous 
section (see also \cite{bh2}) to the case where an additional fast gluon is 
radiated. The interaction of the gluon with the external field is treated 
in the high-energy approximation in analogy to the two quarks. 

\begin{figure}[h]
\begin{center}
\parbox[b]{10cm}{\psfig{width=10cm,file=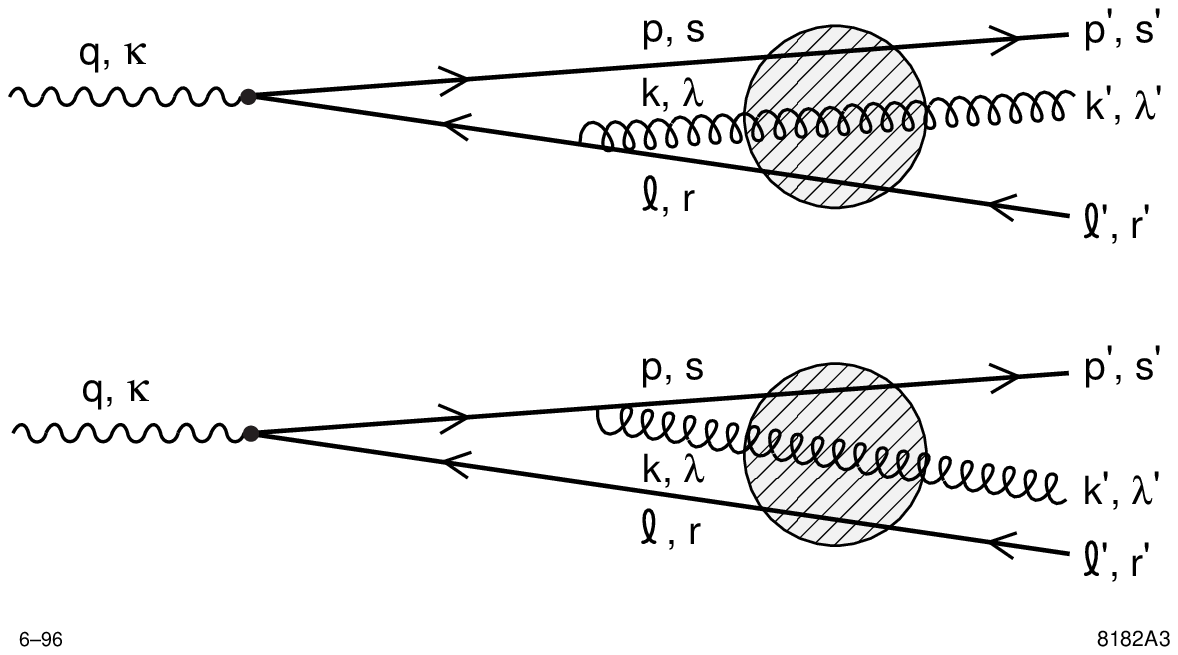}}
\end{center}
\refstepcounter{figure}
\label{dqqg}
{\bf Fig.\ref{dqqg}} Diagrams for the process $\gamma^*\to q\bar{q}g$. 
\end{figure}

The amplitude is given by the sum of the two diagrams shown in 
Fig.~\ref{dqqg}. Separating explicitly the part which describes the 
scattering of the gluon off the external field, the amplitude can be given in 
the form 
\beq
S^b_{\m}=\int\frac{d^4k}{(2\pi)^4} A^a_{\mu\nu}
\frac{-ig^{\nu\rho}}{k^2}B^{ab}_{\rho\s}\e_{(\la')}^{*\s}(k')\, .\label{sbab}
\eeq
Here $A$ refers to the production of the $q\bar{q}g$-system, including the 
interactions of the quarks with the external field, and $B$ describes the 
scattering of the gluon. Like the amplitude $S_\m$ defined in Sect.~3,
$S^b_\m$ is a $3\times 3$ colour matrix. The index $b$ denotes the colour of 
the outgoing gluon. 

Using the approximate $k_-$-independence of $B$, which is analogous to the 
quark scattering amplitude of the previous section, the $k_-$-integration 
in Eq.~(\ref{sbab}) can be performed in such a way that the gluon 
propagator goes on shell. Since $A$ and $B$ are now physical amplitudes and 
therefore gauge invariant, only physical polarizations contribute to the 
gluon propagator connecting $A$ and $B$, 
\beq
S^b_{\m}=\sum_{\lambda=1,2}\int\frac{dk_+d^2k_\perp}{2(2\pi)^3}
A^a_{\mu\nu}\epsilon^{*\nu}_{(\lambda)}(k)\frac{1}
{k_+}\epsilon^\rho_{(\lambda)}(k)B^{ab}_{\rho\s}\e_{(\la')}^{*\s}(k')\, .
\eeq
The expression for the high-energy scattering of the gluon is very similar 
to the analogous expression for the quark given in the previous section, 
\bea
T^g_{\lambda'\lambda}(k,k')&=&
\epsilon^\rho_{(\lambda)}(k)B_{\rho\s}\e_{(\la')}^{*\s}(k')\nn\\
&=&2\pi i\, 2k_+\delta_{\lambda'\lambda}
\delta(k_+'-k_+)(\tilde{F}^\dagger_{\cal A}(k_\perp'-k_\perp)-
(2\pi)^2\de(k'_\t-k_\t))\, .
\eea
A derivation is sketched in Appendix A. The main difference to the quark 
case lies in the eikonal factor, which is now taken in the adjoint 
representation,
\beq
F_{\cal A}^{ab}(x_\perp)\equiv{\cal A}(F(x_\perp))^{ab}\, .
\eeq
The quark propagators $i/\psu$ and $-i/\ls$, where $l=q-p-k$, are treated in 
the high-energy approximation as explained in Sect.~\ref{qq}. The 
$p_-$-integration implicit in $A$ can be performed in such a way that
$p$ goes on shell in the first diagram of Fig.~\ref{dqqg} and $l$ goes on
shell in the second diagram. The $p_+$- and $k_+$-integrations are 
performed using two of the three $\delta$-functions from the amplitudes for 
the scattering off the external field. As a result of these manipulations 
the following expression is obtained, 
\beq
S^a_{\m}=eg\, 2\pi\delta(q_+-p_+'-k_+'-l_+')\int\frac{d^2p_\perp}{(2\pi)^2}
\frac{d^2k_\perp}{(2\pi)^2}\frac{2q_+\cdot {\cal M}_{\m}\cdot C^a}{\left(Q^2+
\frac{p_\perp^2}{1-\alpha-\alpha'}+\frac{k_\perp^2}{\alpha'}+
\frac{l_\perp^2}{\alpha}\right)}\, .\label{saab}
\eeq
Here $\alpha=\l_+/q_+\, ,\,\alpha'=k_+/q_+$ and ${\cal M}_\m$ describes the 
purely partonic part of the amplitude given by 
\beq
{\cal M}_\m=\bar{u}_{s'}(\bar{p})\left[\g_\m\frac{1}{\qs-\bps}
\epsilons_{(\lambda')}(k)-\epsilons_{(\lambda')}(k)\frac{1}{\qs-\bls}
\g_\m\right]v_{r'}(\bar{l})\, .\label{mcal}
\eeq
All the non-abelian eikonal factors are combined in $C^a$, 
\beq
C^a=\int_{x_\t,y_\t,z_\t} e^{i[x_\t(p_\t-p'_\t)+y_\t(k_\t-k'_\t)
     +z_\t(l_\t-l'_\t)]}F(x_\t,y_\t,z_\t)^a\, ,
\eeq
\beq
F(x_\t,y_\t,z_\t)^a={\cal A}(F^\dagger(y_\t))^{ab}
(F^\dagger(x_\t)T^bF(z_\t)) - T^a\, .\label{fxyz}
\eeq
The last term in 
Eq.~(\ref{fxyz}) subtracts the unphysical contribution where none of the 
partons is scattered by the external field (cf. Eq.~(\ref{amp3})). 
One can think of $x_\perp,y_\perp$ and $z_\perp$ as the transverse 
positions at which quark, gluon and antiquark penetrate the proton field,
picking up corresponding non-abelian eikonal factors.

\subsection{Colour structure}\label{cols}

In the phase space region with two of the three final state particles having 
high $p_\perp$, in the $\gamma^*p$ center-of-mass system, 
the form of the three-particle colour 
factor $C^a$ simplifies significantly. To see this, the three possible 
configurations, i.e. high-$p_\perp$ $q\bar{q}$-jets, high-$p_\perp$ 
$qg$-jets and high-$p_\perp$ $\bar{q}g$-jets, have to be distinguished. 

Consider first the case of high-$p_\perp$ quark and antiquark, i.e. 
$p_\perp'^2,\, l_\perp'^2\gg\Lambda^2$. In analogy to the results of 
Sect.~\ref{qq} a leading twist contribution to diffraction can only appear 
if the gluon is relatively soft, i.e. in the region of small $k_\perp'^2$ 
and $\alpha'$. Therefore the relations $k_\t'^2 \sim \La^2$ and 
$\alpha'\ll 1$ are used in the calculations below. This will also be 
justified by the final formulae, which show that these kinematical regions 
dominate the integrations.

The assumption of a smooth external field implies small transverse momentum 
transfer from the proton, i.e. $|p_\t''| \sim \La$, where 
$p_\perp''= p_\perp'-p_\perp$.  The $p_\perp$-integration 
in Eq.~(\ref{saab}) can be trivially replaced by a $p_\perp''$-integration, 
substituting at the same 
time 
\beq
p_\perp=p_\perp'-p_\perp''\qquad\mbox{and}\qquad l_\perp=-p_\perp'+p_\perp''
-k_\perp\, .\label{pl}
\eeq
Neglecting $p_\perp''$ in ${\cal M}$ and in the energy denominator in 
Eq.~(\ref{saab}), which is justified since $|p_\perp''^2|\ll|p_\perp'^2|,\, 
|l_\perp'^2|$, the only remaining $p_\perp''$-dependence is in the colour 
factor $C^a$. This simplifies the $p_\perp''$-integration to
\beq
\int\frac{d^2p_\perp''}{(2\pi)^2}\ C^a\, .
\eeq
Defining $\Delta\equiv p'+k'+l'-p-k-l$ to be the total momentum
transferred from the proton, $C^a$ can be given in the from
\beq
C^a =\int_{x_\t} e^{-ix_\perp\Delta_\perp}\int_{y_\t,z_\t}
e^{i[y_\perp(k_\t-k'_\t)+z_\perp(l_\t-l'_\t)]}F(x_\perp,x_\perp+y_\perp,
x_\perp+z_\perp)^a\, ,
\eeq
where $l_\perp$ is given by Eq.~(\ref{pl}). The $p_\perp''$-integration 
gives a $\delta$-function for the variable $z_\perp$, thus resulting in the 
final formula 
\beq
\int\frac{d^2p_\perp''}{(2\pi)^2}\ C^a = \int_{x_\t}\ e^{-ix_\perp
\Delta_\perp}\int_{y_\t}\ e^{iy_\perp(k_{\perp}-k_{\perp}^{\prime})}
F(x_\perp,x_\perp+y_\perp,x_\perp)^a\, .
\eeq
This result shows that in the kinematical situation with two high-$p_\perp$ 
quark jets and a relatively soft gluon the leading twist contribution is not 
affected by the transverse separation of the quarks. It is the 
transverse separation between quark-pair and gluon which tests large 
distances in the proton field and which can lead to non-perturbative effects.

\begin{figure}[h]
\begin{center}
\parbox[b]{10cm}{\psfig{width=10cm,file=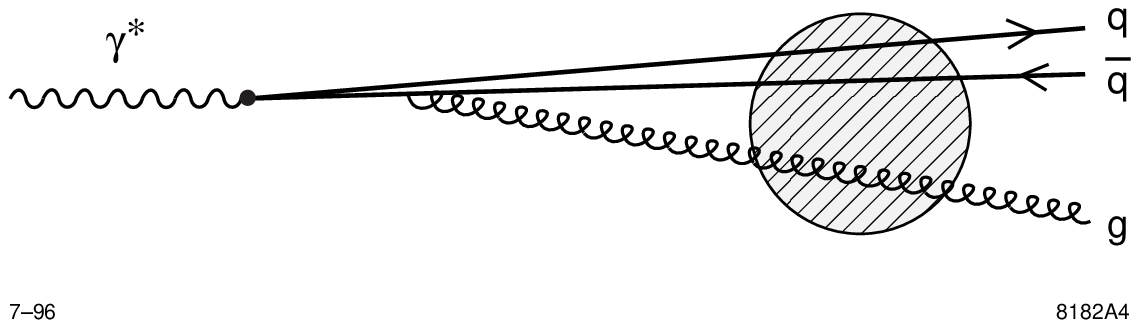}}\\
\end{center}
\refstepcounter{figure}
\label{qqj}
{\bf Fig.\ref{qqj}} Space-time picture in the case of fast, high-$p_\perp$ 
quark and antiquark, passing the proton at small transverse separation
with a relatively soft gluon further away. 
\end{figure}

The colour-singlet projection of the colour-tensor $C$ reads
\beq
S(C)=\2\ \mbox{tr}[C^a T^a]\, .
\eeq
Using the identity 
\beq
{\cal A}(U)^{ab}=2\ \mbox{tr}[U^{-1}T^aUT^b]\quad,\quad U\in \mbox{SU(3)}\, ,
\eeq
the contribution relevant for diffraction, i.e. the production of a 
colour-singlet $q\bar{q}g$-system, becomes
\beq
\int\frac{d^2p_\perp''}{(2\pi)^2}\ S(C)=\int_{x_\t} e^{-ix_\t\Delta_\t}\,
\frac{1}{4}\, \mbox{tr}[\tilde{W}^{\cal A}_{x_\perp}(k_\perp-k_\perp')]\, ,
\label{cf1}
\eeq
\beq
W^{\cal A}_{x_\perp}(y_\perp)={\cal A}(F^\dagger(x_\perp+y_\perp)F(x_\perp))
-1\, .
\eeq
This is analogous to the quark-pair production of the previous section (cf. 
Eq.~(\ref{amp3})). However, now the two lines probing the field 
at positions $x_\perp$ and $x_\perp+y_\perp$ correspond to matrices in the 
adjoint representation. An intuitive explanation of this result is that 
the two high-$p_\perp$ quarks are close together and are rotated in colour 
space like a vector in the octet representation. This situation is 
illustrated in Fig.~\ref{qqj}.

To make this last statement more precise, recall that an upper bound for the 
Ioffe-time of the fluctuation with two high-$p_\perp$ quarks is given by 
$q_0/p_\perp^2$. This means that the distance between the point 
where the virtual photon splits into the $q\bar{q}$-pair and the proton can 
not be larger than $q_0/p_\perp^2$. As long as the pair shares the 
longitudinal momentum of the photon approximately equally, i.e. 
$\a (1-\a) = {\cal O}(1)$, the opening angle is 
$\sim p_\perp/q_0$. Therefore, the transverse distance between quark and 
antiquark is $\sim 1/p_\perp\ll 1/\Lambda$ when they hit the proton. 

In the case where quark and gluon have high transverse momentum 
and the relatively soft, low-$p_\perp$ antiquark is responsible for the 
non-perturbative interaction, we have 
$|p_\perp'^2|\simeq|k_\perp'^2|\gg|l_\perp'^2| \sim \La^2$ 
and $\alpha''\equiv l_+/q_+\ll 1$. 

\begin{figure}[h]
\begin{center}
\parbox[b]{10cm}{\psfig{width=10cm,file=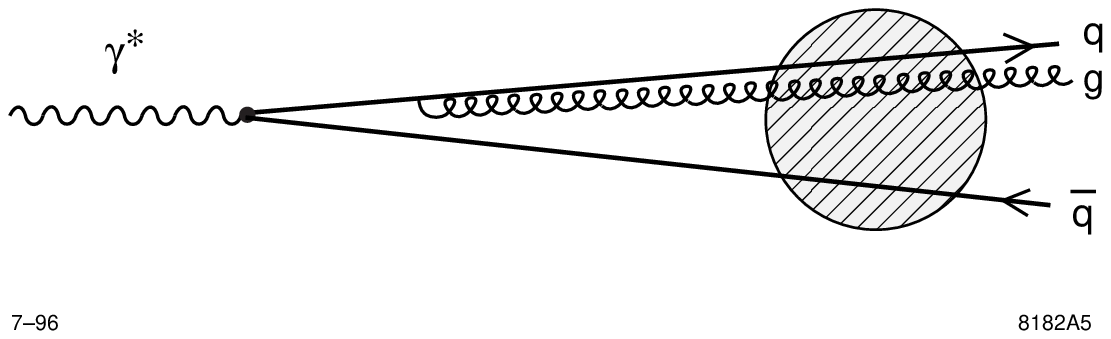}}\\
\end{center}
\refstepcounter{figure}
\label{qgj}
{\bf Fig.\ref{qgj}} Space-time picture in the case of fast, high-$p_\perp$ 
quark and gluon, passing the proton at small transverse separation
with a relatively soft antiquark further away.
\end{figure}

The calculation proceeds along the lines of the soft gluon case described
previously. It is convenient to make the integration over the soft 
transverse momentum explicit by substituting $d^2l_\perp$ for $d^2k_\perp$ 
in Eq.~(\ref{saab}). The $p_\perp$-integration is replaced by a 
$p_\perp''$-integration and the $p_\perp''$-dependence entering ${\cal M}$ 
and the energy-denominator via the relations $p_\perp=p_\perp'-p_\perp''$ 
and $k_\perp=-p_\perp'+p_\perp''-l_\perp$ is neglected. The 
$p_\perp''$-integration gives a $\delta$-function for the variable $y_\perp$, 
leading to the following result for the singlet projection of the colour 
tensor $C$, 
\beq
\int\frac{d^2p_\perp''}{(2\pi)^2}\ S(C)=\int_{x_\t}\ e^{-ix_\perp
\Delta_\perp}\,\frac{2}{3}\,\mbox{tr}[\tilde{W}_{x_\perp}(l_\perp-l_\perp')]
\, .
\eeq
The function $W_{x_\perp}$ has been defined in Eq.~(\ref{wdef}). 
Now quark and gluon, having high transverse momentum, 
are close together and are colour rotated like a vector in the fundamental 
representation (see Fig.~\ref{qgj}). Therefore, the colour structure of the 
amplitude is the same as for the quark-pair production of Sect.~3.

The case of high-$p_\perp$ $\bar{q}g$-jets is completely analogous to the 
case of high-$p_\perp$ $qg$-jets and will not be discussed separately. 

\subsection{Longitudinal cross section}\label{lcs}

In the phase space region of $q\bar{q}$-jets with high transverse momentum 
described at the beginning of the previous subsection the phase space 
integration can be given in the form 
\bea
d\Phi^{(3)}&=&\frac{2}{(2\pi)^9}\, \frac{dp_+'}{2p_+'}d^2p_\perp'
\frac{dk_+'}{2k_+'}d^2k_\perp'\frac{dl_+'}{2l_+'}d^2l_\perp'\delta(q_+-p_+'
-k_+'-l_+')\label{dphi}\\ \nonumber\\&=&\frac{1}{8(2\pi)^8}\frac{Q^2}{q_+x
\alpha'}d\alpha d\alpha'd\xi d^2k_\perp'd^2\Delta_\perp\, .\nonumber
\eea
One obtains analogous expressions for the regions of high transverse 
momentum $qg$-and $\bar{q}g$-jets by replacing  $\alpha'$ 
and $k_\perp'$ by the longitudinal momentum fraction and the
transverse momentum of the soft parton. 

Explicit formulae for the partonic part ${\cal M}$ of the amplitude defined 
in Sect.~\ref{ampl} are given in Appendix C. Treating the colour factors 
as described in the previous subsection explicit expressions for the 
different contributions to the longitudinal diffractive structure function 
are obtained. To stress the common features of these contributions from the 
different phase space regions generic kinematical variables are introduced. 
The longitudinal momentum fraction and the transverse momentum of one of the 
high-$p_\perp$ jets are denoted by $\gamma$ and $r_\perp$ respectively. 
The longitudinal momentum fraction of the relatively soft parton 
is denoted by $\gamma'$, its transverse 
momenta before and after the interaction with the external field are denoted 
by $s_\perp$ and $s_\perp'$. For example, in the case of quark and 
antiquark jets  this notation means that 
$\gamma=\alpha\, ,$ $\gamma'=\alpha'\, ,$ $r_\perp=p_\perp\, ,$ 
$s_\perp=k_\perp\, ,$ and $s_\perp'=k_\perp'\, .$ 

The different contributions to $F_L^D$ can be given in the form 
\beq
F_L^{D,n}(x,Q^2,\xi)=\frac{16\alpha_S}{(2\pi)^5\xi}\ \beta^2(1-\beta)g_L^{(n)}
(\beta)\int_0^1d\gamma\, ,\qquad n=1,2\, .\label{fldi}
\eeq
Here the contribution from the region of high-$p_\perp$ 
$q\bar{q}$-jets is labeled by $n=1$ and the sum of the contributions from 
the regions of high-$p_\perp$ $qg$- and $\bar{q}g$-jets is labeled by 
$n=2$. The trivial $\gamma$-integration has been kept 
explicitly to allow a more detailed description of the final states. 

The $\beta$-spectrum can be different for the two contributions of 
Eq.~(\ref{fldi}). Its dependence on the details of the colour field of the 
proton is given by the two dimensionless functions 
\bea
g_L^{(1)}(\beta)&=&\int_0^\infty\frac{(1+u)^2u\,du}{32(\beta+u)^2}\int
d^2s_\perp'(s_\perp'^2)^2\int_{x_\perp}\left|\int\frac{d^2s_\perp}
{(2\pi)^2}\cdot\frac{\mbox{tr}[\tilde{W}^{\cal A}_{x_\perp}(s_\perp-s_\perp')]
\, t_L^{ij}}{s_\perp'^2(\beta+u)+s_\perp^2(1-\beta)}\right|^2\label{g1}
\\&&\nonumber\\
g_L^{(2)}(\beta)&=&\int_0^\infty\frac{2(1+u)^2du}{9(\beta+u)^2}\int 
d^2s_\perp'(s_\perp'^2)\int_{x_\perp}\left|\int\frac{d^2s_\perp}{(2\pi)^2}
\cdot\frac{s_\perp\cdot\mbox{tr}[\tilde{W}_{x_\perp}(s_\perp-s_\perp')]}
{s_\perp'^2(\beta+u)+s_\perp^2(1-\beta)}\right|^2,\label{g2}
\eea
where the tensor $t^{ij}_L$ is given by
\beq
t_L^{ij}=\delta^{ij}+\frac{2s_\perp^is_\perp^j}{s_\perp'^2}\cdot
\frac{1-\beta}{\beta+u}\, ,\qquad i,j=1,2\, .\label{tij}
\eeq
Note, that in the modulus squared in Eq.~(\ref{g1}) the appropriate 
contraction of the indices of the two tensors $t_L^{ij}$ is assumed.

The integration variable $u$ can be related to the old kinematic variables 
by 
\beq
\gamma'=\frac{s_\perp'^2}{M^2}(1+u)\, .\label{udef}
\eeq
The functions $g^i_L$ have been given in terms of an integral in $u$ to make 
it obvious that they do not depend on any kinematical variable other than 
$\beta$. 

For the transverse momentum $r_\perp$ of the two hard jets the relation
\beq
r_\perp^2=\gamma(1-\gamma)\left(M^2-\frac{s_\perp'^2}{\gamma'}\right)=
\gamma(1-\gamma)M^2\frac{u}{1+u}\label{rp}
\eeq
can be derived. Using this relation and the $\gamma$- and 
$u$-distributions given by Eq.~(\ref{fldi}) and Eqs.~(\ref{g1}),(\ref{g2}) 
the $r_\perp$-distribution of the jets in the final state can be easily 
recovered. 

The above results show that within our model $F_L^D$ has a leading twist 
contribution, suppressed by one power of $\alpha_S$. For $\be$ not too close
to 0 and 1 the integrals in $\gamma$ and $u$ are finite and dominated 
by the region where both $\gamma$ and $u$ are ${\cal O}(1)$. This justifies 
the assumption that $\gamma'\ll 1$ and $s_\perp'^2\ll r_\perp^2$ made at the 
beginning of this section. Eq.~(\ref{rp}) also shows that jets with 
transverse momentum of order $M$ dominate $F_L^D$. This has to be contrasted 
with the case of transverse photon polarization, where the leading 
contribution comes from the production of a $q\bar{q}$-pair with small 
transverse momenta as discussed in Sect.~\ref{qq}.

The details of the $\beta$-spectrum and the $r_\perp$-distribution depend 
on the average over the proton field, which enters via non-abelian eikonal 
factors in the adjoint and fundamental representation (see 
Eqs.~(\ref{g1}),(\ref{g2})). This is a truly non-perturbative effect 
which, in our model, is responsible for leading twist diffraction. It 
is realized by one of the three produced partons, which is slower by a 
factor $\sim \Lambda^2/M^2$ and can develop a large transverse separation 
from the two other partons.

Let us finally consider the behaviour of the diffractive structure function 
at small $\beta$, i.e. at large invariant masses. The $u$-integration in 
$g^{(1)}_L(\beta)$ is divergent for $\beta=0$. The behaviour of $g^{(1)}_L 
(\beta)$ at small $\beta$ is determined by the integration over small values 
of $u$. This behaviour can be obtained assuming $\beta\ll 1$, $u\ll 1$ and 
$u+\beta\ll 1$ in Eq.~(\ref{g1}). In this region the second contribution of
the tensor $t^{ij}_L$ in Eq.~(\ref{tij}) dominates and the following 
expression for $g^{(1)}_L$ is obtained,
\bea
g_L^{(1)}(\beta\to 0)&=&\int_0^\infty\frac{u\,du}{8(\beta+u)^4}\int
d^2s_\perp'\int_{x_\perp}\left|\int\frac{d^2s_\perp}{(2\pi)^2}\cdot
\frac{\mbox{tr}[\tilde{W}^{\cal A}_{x_\perp}(s_\perp-s_\perp')]\ s_\perp^i
s_\perp^j}{s_\perp^2}\right|^2
\\&&\nonumber\\
&=&\frac{1}{48\beta^2}\int d^2s_\perp'\int_{x_\perp}\left|\int
\frac{d^2s_\perp}{(2\pi)^2}\cdot\frac{\mbox{tr}[\tilde{W}^{\cal A}_{x_\perp}
(s_\perp-s_\perp')]\ s_\perp^is_\perp^j}{s_\perp^2}\right|^2\, .
\eea
This means that $F_L^D(x,Q^2,\xi)$ approaches a constant value at fixed 
$\xi$ and $\beta\to 0$. For the inclusive structure function, where
one has to integrate over $\be$, this implies a growth $\sim \ln{(1/x)}$.
Since $g^{(2)}_L$ is less singular than $g^{(1)}_L$ at small $\beta$, the 
region of large diffractive masses is dominated by the configuration with 
high-$p_\perp$ $q\bar{q}$-jets and a relatively soft gluon.

\subsection{Transverse cross section}\label{tcs}

The calculation of the diffractive contribution to the transverse structure 
function proceeds along the lines of the previous subsection, using the 
techniques described in Appendices B and C. However, due to the summation 
over photon polarizations the final formulae are somewhat more complicated. 
The result for the transverse diffractive cross section will be given using 
the generic kinematical variables introduced for the discussion of the 
longitudinal structure function above. Consider the contribution from the 
region of high-$p_\perp$ $q\bar{q}$-jets first, 
\beq
F_T^{D,q\bar{q}}=\frac{4\alpha_S}{(2\pi)^5\xi}\ \beta(1-\beta)^2g_T^{(1)}
(\beta)\int_{\gamma_{min}}^{1-\gamma_{min}}d\gamma\frac{\gamma^2+
(1-\gamma)^2}{\gamma(1-\gamma)}\, .
\eeq
Note that the $\gamma$-integration has a logarithmic divergence in the 
region where quark or antiquark become soft, like the production of 
colour-octet $q\bar{q}$-pairs discussed in Sect.~\ref{qq}. 
Since our calculation is only reliable for $\gamma>\gamma'\sim\Lambda^2/Q^2$, 
the integration leads to a $\ln Q^2/\Lambda^2$ enhancement, 
which was absent in the leading order diffractive structure function given 
in Sect.~\ref{qq}. 

The $\beta$-spectrum is determined by the dimensionless function 
\beq
g_T^{(1)}(\beta)=\int_0^\infty\frac{(1+u)u^2\,du}{16(\beta+u)^2}\int
d^2s_\perp'(s_\perp'^2)^2\int_{x_\perp}\left|\int\frac{d^2s_\perp}
{(2\pi)^2}\cdot\frac{\mbox{tr}[\tilde{W}^{\cal A}_{x_\perp}(s_\perp-s_\perp')]
\, t^{ij}_T}{s_\perp'^2(\beta+u)+s_\perp^2(1-\beta)}\right|^2\, ,\label{g1t}
\eeq
where
\beq
t_T^{ij}=\frac{1}{2}\left(\delta^{ij}+\frac{2s_\perp^is_\perp^j}{s_\perp'^2}
\cdot\frac{1-\beta}{\beta+u}\right)\sqrt{1+\left(\frac{\beta(1+u)}
{u(1-\beta)}\right)^2}.  
\eeq
Contraction of the transverse tensor indices is 
assumed in Eq.~(\ref{g1t}).

It can be shown, following the discussion at the end of the last subsection, 
that $g_T^{(1)}(\beta)\sim 1/\beta$ at $\beta\to 0$, resulting in a constant 
behaviour of the diffractive structure function in this region. The same
effect has previously been observed in \cite{bawu}, where the diffractive
interaction with the proton is treated by means of 2-gluon exchange.

The sum of the contributions from the region of high-$p_\perp$ quark-gluon
or antiquark-gluon jets is given by
\beq
F_T^{D,qg}=\frac{4\alpha_S}{(2\pi)^5\xi}\ \beta(1-\beta)^2\!\!\!\!
\int\limits_{\gamma_{min}}^{1-\gamma_{min}}\!\!\!\!d\gamma\left[g_T^{(2)}
(\beta)+\frac{1}{\gamma(1\!-\!\gamma)}g_T^{(3)}(\beta)+\frac{\gamma^3+(1\!-
\!\gamma)^3}{\gamma(1\!-\!\gamma)}g_T^{(4)}(\beta)\right]\, ,
\eeq
where for $m=2,3,4$ the functions $g_T^{(m)}(\beta)$ are defined by
\beq
g_T^{(m)}(\beta)=\int_0^\infty\frac{2(1+u)u\,du}{9(\beta+u)^2}\int
d^2s_\perp'(s_\perp'^2)\int_{x_\perp}\left|\int\frac{d^2s_\perp}
{(2\pi)^2}\cdot\frac{s_\perp\,\mbox{tr}[\tilde{W}_{x_\perp}(s_\perp-s_\perp')]
\, t^{(m)}}{s_\perp'^2(\beta+u)+s_\perp^2(1-\beta)}\right|^2\, ,
\eeq
\beq
t^{(2)}=1\quad,\quad t^{(3)}=-\frac{\beta+u}{u(1-\beta)}\quad,\quad 
t^{(4)}=t^{(2)}+t^{(3)}\, .
\eeq
As in the longitudinal case, the contribution from the region with a 
high-$p_\perp$ gluon-jet is negligible at small $\beta$.

The results of the present section are the above explicit expressions for 
hard radiative corrections to the diffractive structure functions, 
calculated in the semiclassical approach of \cite{bh2}. These radiative
corrections yield high-$p_\t$ jets, whereas the leading contribution is
kinematically dominated by the aligned-jet configuration with small
transverse momentum. This is similar to expectations based on the 
QCD-improved aligned-jet model \cite{fs}.

Final states with high-$p_\t$ jets in diffractive electroproduction have
recently been also considered in \cite{diehl,balo}. The 
main difference to our approach is the assumption of two-gluon exchange in 
these calculations. This leads to a different $x$-dependence of the 
jet cross sections.

\section{Conclusions}

It has been the purpose of this paper to investigate the applicability of the 
semiclassical approach to diffraction, to improve its technical 
implementation, and to extend it to the case of an additional final 
state gluon, radiated by the produced $q\bar{q}$-pair. 

As has been shown in Sect.~\ref{sc}, the situation where the proton is 
scattered elastically can be described as quark-pair production in a 
given external colour field, averaging over all field configurations of the 
proton on the amplitude level. This averaging procedure has formally been 
defined using the Schr\"odinger wave functional of the proton. 

The calculation of colour-singlet and colour-octet quark pair production in 
an external colour field had already been performed in \cite{bh2}. In the 
present paper this calculation has been technically improved, thereby 
establishing its similarity to well-known light-cone techniques. This 
simpler technique made it possible to consider ${\cal O}(\a_S)$-corrections 
arising from an additional fast gluon in the final state. The cross sections 
for diffractive final states with a quark pair and one gluon have been given 
in the phase space region where two of the partons have large transverse 
momentum.

{}From these calculations an intuitive picture of leading twist diffraction 
emerges, which is very similar to Bjorken's aligned jet model. The leading 
contribution to both diffractive and non-diffractive deep inelastic events 
at small $x$ 
is given by the process where the incoming virtual photon splits into a 
quark-pair long before the proton. When the quarks travel through the 
proton field they pick up non-abelian eikonal factors associated with their 
trajectory. These eikonal factors are sufficient to provide the small 
momentum transfer required to make the final state real. 

If the relative transverse momentum of the quarks is large, the quarks have 
a small transverse separation when they travel through the proton field. 
Expanding in the transverse distance of the quarks it becomes obvious that 
the process is essentially perturbative and equivalent to one-gluon 
exchange. A constant contribution to $F_L$ and a $\ln Q^2$ enhanced 
contribution to $F_2$ arise. No leading twist diffraction is possible. 

If, however, the transverse momenta of the quarks are small, their 
trajectories pass the proton field at large transverse distance. The two 
eikonal factors test large transverse distances inside the proton making the 
process non-perturbative. The result is no longer equivalent to an exchange 
of a finite number of gluons. This gives a leading twist contribution to 
$F_2$ and $F_2^D$, the latter being interpreted as the fraction of 
events where the produced quark pair is in a colour-singlet state. Neither a
$\ln Q^2$ enhancement of $F_2^D$ nor a leading twist contribution to 
$F_L^D$ are found. 

The picture becomes more complicated when the radiation of an additional 
gluon is included. Now three eikonal factors test the proton field, 
corresponding to the trajectories of the two quarks and the gluon. The 
gluonic eikonal factor is in the adjoint representation. Leading twist 
diffractive 
processes appear when at least one of the three partons has small transverse 
momentum and carries a small fraction of the longitudinal momentum of the 
photon. The two other partons can have large transverse momentum. In this
kinematical situation the two high-$p_\perp$ partons pass the field close 
together, acting effectively as one particle. Therefore, only two Wilson 
lines test the proton field, analogously to the pure $q\bar{q}$-case. The 
corresponding colour rotation matrix is in the fundamental representation if 
the gluon combines with quark or antiquark, and in the adjoint 
representation if the two quarks combine to form a colour octet state.

The colour singlet projection of the $q\bar{q}g$-final state is interpreted 
as the diffractive contribution. As a result, leading twist, 
$\alpha_S$-suppressed contributions to $F_2^D$ and $F_L^D$ are obtained. In 
the longitudinal case 
this is the leading diffractive process. It is dominated by configurations 
containing two jets with high transverse momentum, $|p_\perp|\sim M$, 
and is free from infrared divergences. 
No $\ln Q^2$ term appears. In the case of 
transverse polarization the diffractive cross section also contains a 
leading twist contribution from high-$p_\perp$ jets. 
However, now an infrared divergence appears in the 
region where the transverse momentum becomes small, leading to an 
$\alpha_S\ln Q^2$-term in the diffractive structure function. 

Several qualitative phenomenological predictions can be derived from this 
picture of diffractive processes. Notice first, that the scaling violation 
associated with the $\ln Q^2$-terms appears in diffraction only in 
next-to-leading order. This means, that one would expect the ratio 
$F_2^D/F_2$ to decrease approximately like $\ln Q^2$. Also high-$p_\perp$ 
jets and diffraction of the longitudinally polarized photon appear only 
at order $\alpha_S$. Therefore, in the $\gamma^*p$ center-of-mass system 
less high-$p_\perp$ jets should be visible in diffractive processes 
than in inclusive processes. It is, however, interesting to observe that 
$F_L^D$ is, unlike $F_2^D$, dominated by high-$p_\perp$ jets.

The details of the $\beta$-spectrum and of the $p_\perp$-distribution of 
jets in diffractive processes depend on the average over the proton field 
configurations appearing in the semiclassical approach. They are given in 
terms of integrals over non-abelian eikonal factors testing the proton 
field. At small $\be$ the structure functions approach constants different 
from zero. In this region they are dominated by contributions from 
high-$p_\t$ quark jets accompanied by a relatively soft gluon with small 
transverse momentum.

An important aspect of the semiclassical approach, which requires further 
study, is the energy dependence of the obtained cross sections. To leading
order, the scattering in a classical background field yields a 
flat behaviour at small $x$, i.e., $F_2 \sim$ const. and $F_2^D \sim
1/\xi$.  This corresponds to a classical bremsstrahl spectrum of
gluons, $xg(x) =$const. The difference by one unit
in the exponents of inclusive and diffractive structure functions is in 
agreement with our previous results \cite{bh1,bh2,b0}, but certainly a 
stronger increase is observed individually in both cases. Two potential 
sources for such an increase are loop corrections to the partonic part of 
our amplitude and radiation of additional gluons.
\\*[.0cm]

We would like to thank J.~Bartels, M.~Beneke, S.J.~Brodsky and L.~Frankfurt 
for valuable discussions and comments. A.H. has been supported by the Feodor 
Lynen Program of the Alexander von Humboldt Foundation.

\newpage
\section*{Appendix A}

In the following we shall give a brief derivation of the quark and gluon
matrix elements used in Sect.~3 and Sect.~4.1.

Consider a fast moving particle with momentum $l^{\prime}$,
where
$l_+^{\prime}$
is
the
large
component. The wave function of an antiquark propagating through a colour 
field $G^a_{\mu}(x)$ satisfies the Dirac equation 
\beq
(i \Ds - m)\ \psi_v(x) = 0\quad,\qquad D_{\mu}=\partial_{\mu}+iG_{\mu}\ .
\eeq
An approximate solution for a fast antiquark in a `soft' colour field, which 
approaches a plane wave solution with momentum $l^{\prime}$ of the free 
Dirac equation as $x^0 \rightarrow \infty$, is given by
\beq\label{wv}
\psi_v(x) = e^{i l^{\prime} x}\ V_0(x)\ v(l^{\prime})\ ,
\eeq
where $\ls^{\prime} v(l^{\prime}) = 0$ and
\beq
V_0(x) = P \exp{\left({i\over 2}\int_{x_+}^{\infty} dx_+'
                G_-(x'_+,x_-,x_{\perp})\right)}\ .
\eeq
Here colour indices of the spinor $v$ and the matrix $V_0$ have been
dropped. Corrections to this approximate solution are suppressed by powers 
of $1/l^{\prime}_+$. Similarly, one obtains for a fast moving quark,
\beq\label{wu}
\bar{\psi}_u(x) = \bar{u}(l^{\prime})\ U_0(x)\ e^{i l^{\prime} x}\ ,
\eeq
where $\bar{u}(l^{\prime})\ls^{\prime} =0$ and
\beq
U_0(x) = P \exp{\left({i\over 2}\int_{\infty}^{x_+} dx'_+
                 G_-(x'_+,x_-,x_{\perp})\right)}\ =\ V_0^{\dagger}(x)\ .
\eeq
The non-abelian phase factors satisfy the differential equations
\beq\label{deqvu}
l^{\prime}\cdot D\ V_0\ =\ U_0\ l^{\prime}\cdot \Dl\ =\ 0\ ,
\eeq
where $\Dl_{\mu}=\pdl - iG_{\mu}$.

The matrix element for the antiquark needed in Sect.~3 is now easily
evaluated. Using Eq.~(\ref{deqvu}) and $l'\simeq l$, $\bar{v}_r(l)
\gamma^{\mu} v_{r'}(l')\simeq 2\ l^\m\delta_{rr'}$, one obtains
\bea\label{tv}
T_{r,r'}(l,l') &=& \int d^4y\ \bar{v}_r(l) e^{-ily}\Gs(y) \psi_v(y)\nn\\
 &=& \int d^4y\ e^{i(l'-l)y} \bar{v}_r(l)\Gs (y)\ V_0(y)\ v_{r'}(l')\nn\\
 &\simeq&  2 i l_+ \delta_{rr'} \int d^4y\ e^{i(l'-l)y} 
                  {\partial\over \partial y_+} V_0(y)\ .
\eea
In Sections 3 and 4 we have considered diffractive masses $M = {\cal O}(Q)$.
This implies for the momentum transfer, $l'_- - l_- \ll \Lambda$.
Assuming, as discussed in Sect.~2, that the proton is localized at 
$y_3 \approx 0$, i.e. $y_+ = y_- + {\cal O}(1/\Lambda)$, we can then write
\bea
\int dy_+\ e^{i(l^{\prime}_- -l_-)y_+/2}\ {\partial\over \partial y_+}
  V_0(y)
&\simeq& e^{i(l'_- - l_-)y_-/2}\ \int dy_+\ {\partial\over \partial y_+}
          \ V_0(y)\nn\\
&=& -\left(V_0(-\infty,y_-,y_{\perp}) - 1\right)\ .
\eea
Inserting this in Eq.~(\ref{tv}) yields the final result
\beq
T_{rr'}(l,l') =  -2\pi i\ 2 l_+ \delta_{rr'} \delta(l_+-l_+') 
         \left(\tilde{F}(l_{\perp}-l'_{\perp}) - 
               (2\pi)^2 \delta^2(l_{\perp}-l'_{\perp})\right)\ ,
\eeq
where we have used $l_+ \delta(l_+-l'_+)\simeq l_0 \delta(l_0-l'_0)$, and
\bea
\tilde{F}(l_{\perp}-l'_{\perp}) &=&
\int_{y_{\perp}}\ e^{i(l_{\perp}-l'_{\perp})y_{\perp}} F(y_{\perp})\ , \nn\\
F(y_{\perp}) &=& P \exp{\left({i\over 2} \int_{-\infty}^{\infty} dy_+ 
G_-(y_+,y_-,y_{\perp})\right)}\ .
\eea

A fast moving gluon can be treated similarly. The equation of motion for 
the gluon field reads,
\beq
\partial_{\mu} F_{\mu\nu} + i [A_{\mu},F^{\mu\nu}] = 0\ .
\eeq
Expanding $A_\m$ around the classical background field, $A_{\mu} \rightarrow
G_{\mu} + A_{\mu}$, one obtains for a transverse gluon in the high energy 
approximation, where terms of order $1/l^{\prime}_+$ are dropped, the wave 
equation 
\beq \label{wglue}
(\partial^2 - 2\ l^{\prime}\cdot G^{\cal A}) A_{\mu} = 0\ .
\eeq
Here $G^{\cal A}$ denotes the background field in the adjoint representation,
\beq
G^{\cal A}_{\mu} = G^a_{\mu} T^a\quad,\qquad T^a_{bc}=-i f_{abc}\ ,
\eeq
where $f_{abc}$ are the SU(3) structure constants. A solution of 
Eq.~(\ref{wglue}) up to terms ${\cal O}(1/l^{\prime}_+)$, which approaches a 
plane wave with momentum $l^{\prime}$ as $x^0 \rightarrow \infty$, is given 
by 
\beq\label{gwf}
A_{\mu}(x) = e^{i l^{\prime} x}\ V^{\cal A}_0(x)\ \e^*_{\mu}(l^{\prime})\ ,
\eeq
where $l'\cdot \e^*(l') =0$ and
\beq
V^{\cal A}_0(x) = {\cal A}(V_0(x)) 
                = P \exp{\left({i\over 2}\int_{x_+}^{\infty} dx_+'
                  G^{\cal A}_-(x'_+,x_-,x_{\perp})\right)}\ .
\eeq
Obviously, $V_0^{\cal A}$ satisfies the equation
\beq
l^{\prime}\cdot D^{\cal A}\ V_0^{\cal A} = 0\ .
\eeq

The matrix element for the gluon, needed in Sect.~4.1, is now easily 
calculated. Inserting in the expression
\beq
T^g_{\lambda',\lambda} = \int d^4z\ A^\m_{(\lambda')}(z)\ 2l\cdot G(z) 
\e_{(\la)\m}(l)e^{-ilz}
\eeq
the gluon wave function (\ref{gwf}) yields the transition matrix
element
\beq
T^g_{\lambda'\lambda}(l^{\prime},l)=2\pi i\, 2l_+\delta_{\lambda'\lambda}
\delta(l^{\prime}_+- l_+)\left(\tilde{F}^\dagger_{\cal A}(l^{\prime}_\perp-
l_\perp)-(2\pi)^2\delta^2(l^{\prime}_\perp-l_\perp)\right)\, .
\eeq

\section*{Appendix B}

The computation of longitudinal and transverse structure functions in
Sect.~3 requires the evaluation of the product of an amplitude and its
complex conjugate which depend on different transverse momenta, respectively.
To evaluate these products, the following representation of spinors is 
useful (cf.~\cite{itz}),
\beq
u_r(p) = {\psu + m\over \sqrt{p_0 + m}}\ \ph_r(m,0)\ ,\ 
v_s(l) = {-\ls + m\over \sqrt{l_0 + m}}\ \chi_s(m,0)\ .
\eeq
This immediately yields,
\bea
u(\tilde{p})\bar{u}(p) 
 &=& {1\over \sqrt{(p_0 + m)(\tilde{p}_0 + m)}} (\pts + m)
                    {1+\g_0\over 2}(\psu + m)\ ,\quad \\
v(l)\bar{v}(\tilde{l}) 
 &=& -{1\over \sqrt{(l_0 + m)(\tilde{l}_0 + m)}} (\ls - m)
                    {1-\g_0\over 2} (\lts - m)\ ,
\eea
where the sum over spins is understood. In the high energy expansion,
and restricting ourselves to the massless case $m=0$, one obtains
\bea
u(\tilde{p})\bar{u}(p)
 &=& \2 p_+\g_- + \2 p_-\g_0\g_+ + \2 \tilde{p}_-\g_0\g_-
    + p_\t\g_\t\ \g_- {1-\g_0\over 2} \nn\\ 
 && - \tilde{p}_\t\g_\t\ {1+\g_0\over 2} \g_-
    + {2\over p_+}\ \tilde{p}_\t\g_\t\ p_\t\g_\t {1-\g_0\over 2}\ 
    +\ {\cal O}\left({1\over p_+^2}\right)\ ,\\
v(l)\bar{v}(\tilde{l})
 &=& \2 l_+\g_- - \2 \tilde{l}_-\g_0\g_+ - \2 l_-\g_0\g_- 
    - \tilde{l}_\t\g_\t\ \g_- {1+\g_0\over 2} \nn\\ 
 && + l_\t\g_\t\ {1-\g_0\over 2} \g_-
    - {2\over l_+}\ l_\t\g_\t\ \tilde{l}_\t\g_\t {1+\g_0\over 2}\ 
    +\ {\cal O}\left({1\over l_+^2}\right)\ ,
\eea
where we have used $\tilde{p}_+=p_+,\ \tilde{l}_+=l_+$. It is now 
straightforward to
evaluate the matrix elements needed for the longitudinal and the transverse
structure functions. The results are, to leading order in the large momenta
$p_+$ and $l_+$ ($i=1,2$),
\bea
\mbox{tr}\left[u(\tilde{p})\bar{u}(p) \g^0 v(l)\bar{v}(\tilde{l})\g^0\right]
&=& 2\ p_+ l_+ \label{ml}\\
\mbox{tr}\left[u(\tilde{p})\bar{u}(p) \g^i v(l)\bar{v}(\tilde{l})\g^i\right]
&=& 4\ {\a^2 + (1-\a)^2 \over \a (1-\a)}\ p_\t \tilde{p}_\t \label{mt}\ .
\eea

The differential cross section for quark pair production reads
\beq
d\s_{\m\n} = {2\pi\over q_+}\ T^*_\m T_\n\ d\PH^{(2)}\ ,
\eeq
where $T_\m$ is the amplitude given in Eq.~(\ref{amp3}), and $d\PH^{(2)}$ is 
the phase space volume,
\beq\label{phs}
d\PH^{(2)} = {2\over (2\pi)^6}\ {dp_+'\over 2p_+'}d^2p'_\t {dl_+'\over 2l_+'}
       d^2l_\t'\ \de (q_+-p_+'-l_+')\ .
\eeq
Changing variables,
\beq
p'_\t = \D_\t - l'_\t\ ,\quad l'_+ = \a q_+\ ,\quad
\xi = x\ {Q^2 + M^2\over Q^2}\ ,
\eeq
using ${l'_\t}^2 \simeq \a(1-\a) M^2$, and integrating over the orientation
of $l'_\t$, one obtains
\beq
d\PH^{(2)}={1\over 4(2\pi)^5}{1\over q_+}{Q^2\over x} d\a d\xi\ d^2\D_\t\ .
\eeq

Eqs.~(\ref{amp3}), (\ref{ml}) and (\ref{phs}) yield for the longitudinal
structure function
\bea
dF_L &=& {Q^2\over \pi e^2}\ d\s_L\ 
        \simeq\ {4 Q^4\over \pi e^2 q_+^2}\ d\s_{00}\\
     &=& {4\ Q^6\over (2\pi)^7 \beta}\ {d\xi\over \xi}\ d\alpha\ 
         (\alpha(1-\alpha))^3\
         \int_{x_{\perp}} \left|\int d^2 l_{\perp} 
         \frac{\tilde{W}_{x_{\perp}}(l_{\perp}-l'_{\perp})}{N^2 + l_{\perp}^2}
         \right|^2\ .
\eea
Here the integration over the transverse momentum $\D_\t$ has been carried
out. The integrand can be expanded around $l=l'$. Shifting the integration
variable $l_\t$ to $k_\t=l_\t-l'_\t$, the Taylor expansion in powers of 
$k_\t$ of the denominator yields
\beq
{1\over N^2 + (k_\t + l'_\t)^2} = {1\over N^2 + {l'}_\t^2}
   - {2k_\t l'_\t \over (N^2 + {l'}_\t^2)^2}\ +\ \ldots\ .
\eeq
{}From the definition of the colour matrix $W_{x_\t}(y_\t)$ in
 Eq.~(\ref{wdef}) it is clear that
\bea
\int d^2k_\t\ \tilde{W}_{x_\t}(k_\t) &=& W_{x_\perp}(0)\,\,=\,\,0\ ,\nn\\ 
\int d^2k_\t\ k_\t \tilde{W}_{x_\t}(k_\t) &=& -i (2\pi)^2\ 
       \partial_\t W_{x_\t}(0)\ .
\eea
Using rotational invariance, i.e. 
$l'_i l'_j \rightarrow \2 \de_{ij}\ {l'}^2_\t$, and the relations
\beq
{l'}^2_\t \simeq \a(1-\a){1-\be\over \be} Q^2\ ,\quad
N^2 + {l'}^2_\t \simeq \a(1-\a){Q^2\over \be}\ ,
\eeq
one obtains the final result (\ref{fl}).

The transverse structure function can be evaluated in a similar way.
Summing over the transverse polarizations and using the matrix element
(\ref{mt}), one obtains
\bea\label{ftb}
dF_T &=& 
{Q^2\over \pi e^2}\ \2 \sum_{\la=1,2} \e_{\m}^{*(\la)}\e_{\n}^{(\la)}
          \ d\s^{\m\n}
       =  {Q^2\over \pi e^2}\ \2\ d\s^{ii}\nn\\
      &=& { Q^4\over (2\pi)^7 \beta}\ {d\xi\over \xi}\ d\alpha\ 
          \alpha(1-\alpha)(\a^2 + (1-\a)^2)
         \int_{x_{\perp}} \left|\int d^2 l_\t 
         \frac{l_\t \tilde{W}_{x_\t}(l_\t-l'_\t)}{N^2 + l_\t^2}
         \right|^2\!\!.
\label{dft}
\eea
Performing again a Taylor expansion in $k_\t=l_\t-l'_\t$ and using rotational
invariance, one obtains the expression (\ref{ft0}), which is similar to
(\ref{fl}). The expansion is valid for $\a > \La^2/Q^2$. Integration over
$\a$ in the range $\La^2/Q^2$ to 1/2 yields the factor $\ln{Q^2/\La^2} -1$.

In contrast to $F_L$ there is an additional contribution to $F_T$, which is
most easily discussed in the case of diffraction, where
\beq
W_{x_{\perp}}(y_{\perp}) \rightarrow {1\over \sqrt{3}}\ 
          \mbox{tr}[W_{x_{\perp}}(y_{\perp})]\ .\nn
\eeq
Because of $\partial_\t \mbox{tr}[W_{x_\t}(0)] = 0$,
only the region of small $\a$ yields a leading twist contribution to $F_2^D$.
Defining a new integration variable $\r$ by
\beq
\a = {\be\over Q^2}\ \r^2\ ,
\eeq
and using $1 - \a \simeq 1$,
\beq
N^2 \simeq \be\ \r^2\ ,\quad |l'_\t| \simeq \r\ \sqrt{1-\be}\ ,
\eeq
one obtains the expression (\ref{f2d1}),(\ref{f2d2}) for $F_2^D$. Since in 
the integral for $F_2^D$ large values of $\a$ do not contribute at leading 
twist, the integration over $\r$ can be extended to $\infty$. The complete 
expression for the inclusive structure function $F_T$, including the 
function $f(\beta)$ of Eq.~(\ref{ft}), can be obtained analogously, but 
without taking the trace over $W_{x_\perp} (y_\perp)$.

\section*{Appendix C}

Here we give some technical details of the calculation leading to the 
diffractive contributions of Sections \ref{lcs} and \ref{tcs}. 
The main problem is the calculation of the partonic part ${\cal M}$ of the 
amplitude given by Eq.~(\ref{saab}). In the following, the necessary 
manipulations will be described for the longitudinal cross section in the 
case of high-$p_\perp$ quark-antiquark jets. A complete derivation of
$F_L^{D,1}$, Eq.~(\ref{fldi}), is obtained. For the other diffractive
contributions only the results for the corresponding partonic parts 
${\cal M}$ will be given.

The longitudinal cross section can be calculated using the relation 
$\sigma_L=(Q^2/q_0^2)\sigma_{00}$. Therefore it is sufficient to calculate 
${\cal M}$, given by Eq.~(\ref{mcal}), for the unphysical photon 
polarization $\epsilon(q)=(1,\vec{0})$, 
\beq
{\cal M}_0=\bar{u}_{s'}(\bar{p})\left[\gamma_0\frac{\qs-\bps}{(q-\bar{p})^2}
\epsilons-\epsilons\frac{\qs-\bls}{(q-\bar{l})^2}\gamma_0\right]v_{r'}
(\bar{l})\, .\label{ma1}
\eeq
Here $\epsilon_{\lambda'}(k)=\epsilon$ has been used for brevity. The 
high-energy approximation for the quark propagator introduced in 
Sect.~\ref{qq} (see in particular Eqs. (\ref{prop1}),(\ref{prop2})) 
corresponds to the replacements 
\bea
\qs-\bps&\simeq&\bls+\ks=\sum_{\rho}v_\rho(\bar{l})\bar{v}_\rho(\bar{l})+\ks
\\ \nonumber\\ \qs-\bls&\simeq&\bps+\ks=\sum_{\rho}u_\rho(\bar{p})
\bar{u}_\rho(\bar{p})+\ks\, .
\eea
Note, that $k$ is an on-shell vector. Furthermore, it can be shown that the 
terms proportional to $\ks$ are suppressed in Eq.~(\ref{ma1}) in the limit 
$k_\perp^2\ll p_\perp^2\simeq l_\perp^2$, $\alpha' \ll 1$, which corresponds 
to the case of high-$p_\perp$ $q\bar{q}$-jets. Making use of the relations
\beq
\bar{u}_{s'}(\bar{p})\epsilons u_\rho(\bar{p})=(2\bar{p}\epsilon)
\delta_{s'\rho}\quad,\quad\bar{v}_\rho(\bar{l})\epsilons v_{r'}(\bar{l})=
(2\bar{l}\epsilon)\delta_{\rho r'}
\eeq
the following expression for ${\cal M}_0$ can be derived,
\beq
{\cal M}_{0,q\bar{q}}=2\bar{u}_{s'}(\bar{p})\gamma_0v_{r'}(\bar{l})\left(
\frac{\bar{l}\epsilon}{(q-\bar{p})^2}-\frac{\bar{p}\epsilon}{(q-\bar{l})^2}
\right)\, .\label{ma2}
\eeq
Here the index $q\bar{q}$ specifies the considered kinematical region of
high-$p_\perp$ quark and antiquark.

With $\epsilon_-=2\epsilon_\perp k_\perp/k_+$, which follows from 
$\epsilon k=0$, the relations
\beq
\bar{l}\epsilon=\frac{\alpha}{\alpha'}\epsilon_\perp k_\perp
-\epsilon_\perp l_\perp
\quad,\quad \bar{p}\epsilon=\frac{1-\alpha}{\alpha'}\epsilon_\perp k_\perp-
\epsilon_\perp p_\perp\label{lpe}
\eeq
are obtained. The denominators in Eq.~(\ref{ma2}) can be given in the form 
\beq
(q-\bar{p})^2\simeq-\alpha N^2\quad,\quad (q-\bar{l})^2\simeq-(1-\alpha)N^2
\left(1+\frac{2k_\perp p_\perp}{\alpha(1-\alpha)N^2}+{\cal O}(k_\perp^2)
\right)\, ,\label{den}
\eeq
where
\beq
N^2=Q^2+\frac{p_\perp^2}{\alpha(1-\alpha)}\, .
\eeq
Although the term $\sim k_\perp$ in Eq.~(\ref{den}) is small compared to the 
leading term, it can not be neglected. In combination with the term 
$\sim k_\perp/\alpha'$ of Eq.~(\ref{lpe}) it will give a finite contribution 
of order $k_\perp^2/\alpha'$. 

Inserting (\ref{lpe}) and (\ref{den}) into Eq.~(\ref{ma2}) and keeping only 
the leading contributions the following formula for ${\cal M}_{0,q\bar{q}}$ 
is obtained, 
\beq
{\cal M}_{0,q\bar{q}}=\frac{2\bar{u}_{s'}(\bar{p})\gamma_0v_{r'}(\bar{l})}{
\alpha(1-\alpha)N^2}\left(\epsilon_\perp p_\perp+\frac{(2\epsilon_\perp 
k_\perp)(p_\perp k_\perp)}{\alpha'N^2}\right)\, .
\eeq

Notice, that ${\cal M}_{0,q\bar{q}}$ depends on the intermediate gluon 
momentum $k$ and 
the integration over this variable has to be performed independently for 
the amplitude and its complex conjugate. Therefore, we do actually not need 
the square of ${\cal M}_{0,q\bar{q}}$ but the product 
${\cal M}_{0,q\bar{q}}^*(k){\cal M}_{0,q\bar{q}}(\tilde{k})$. Here $k$ and 
$\tilde{k}$ are two independent integration 
variables. Summing over transverse gluon polarizations and quark and 
antiquark helicities the following result is derived,
\beq
\sum_{\lambda's'r'}{\cal M}_{0,q\bar{q}}^*(k){\cal M}_{0,q\bar{q}}(\tilde{k})
=\frac{8q_+^2}{\alpha
(1-\alpha)N^4}\left(p_\perp+k_\perp\frac{(2p_\perp k_\perp)}{\alpha'N^2}
\right)\left(p_\perp+\tilde{k}_\perp\frac{(2p_\perp \tilde{k}_\perp)}{\alpha'
N^2}\right)\, .
\eeq
Using this expression and the colour factor of Eq.~(\ref{cf1}) together with 
the amplitude (\ref{saab}) and the phase space formula (\ref{dphi}) the 
cross section $\sigma_L$ can be calculated. Note, that in Eqs.~(\ref{g1}),
(\ref{g2}) the phase space integration over $\alpha'=\gamma'$ has been 
substituted by an integration over $u$, defined in Eq.~(\ref{udef}).

For the other possible kinematical configurations in both the longitudinal 
and the transverse case similar techniques can be used to perform the 
calculation of ${\cal M}$. In the following we simply state the results for 
the polarization sums over ${\cal M}^*(k){\cal M}(\tilde{k})$. The 
kinematical variables are the same as in Sect.~\ref{qqg}.

For the longitudinal cross section in the case of high-$p_\perp$ quark and
gluon jets the following polarization sum is required,
\beq
\sum_{\lambda's'r'}{\cal M}^*_{0,qg}(l){\cal M}_{0,qg}(\tilde{l})=
\frac{4(1-\alpha)q_+^2}{N^4}\frac{l_\perp\tilde{l}_\perp}{\alpha''}\, .
\eeq
To derive $F_L^{D,2}$, Eq.~(\ref{fldi}), the contribution from 
high-$p_\perp$ antiquark-gluon jets has to be added, which is proportional
to $\alpha$ and renders the cross section independent of the momentum 
fraction.

The diffractive contribution to the transverse structure function from the 
region of high-$p_\perp$ quark and antiquark jets can be calculated from 
\bea
\frac{1}{2}\sum_{i=1,2}\sum_{\lambda's'r'}{\cal M}^*_{i,q\bar{q}}(k)
{\cal M}_{i,q\bar{q}}(\tilde{k})  =  \frac{8(\alpha^2+(1-\alpha)^2)}{\alpha(1-
\alpha)N^4} \, \times  
\eea
\bea
\left(\!\frac{p_\perp^ip_\perp^j}{\!\alpha(\!1\!-\!\alpha\!)}\!-\!\frac{k_\perp^i
k_\perp^j}{\alpha'}\!+\!\frac{(2p_\perp k_\perp)p_\perp^ik_\perp^j}{\!\alpha(\!1\!-\!\alpha\!)N^2\alpha'}\!-\!\frac{N^2}{2}\!\delta^{ij}\!\right)\!
\left(\!\frac{p_\perp^ip_\perp^j}{\alpha(\!1\!-\!\alpha\!)}\!-\!\frac{\tilde{k}_\perp^i\tilde{k}_\perp^j}{\alpha'}\!+\!\frac{(2p_\perp \tilde{k}_\perp)p_\perp^i\tilde{k}_\perp^j}{\alpha(\!1-\!\alpha\!)N^2\alpha'}\!
-\!\frac{N^2}{2}\!\delta^{ij}\!\right) . \nonumber 
\eea
In the case of high-$p_\perp$ quark and gluon the corresponding contribution 
reads
\bea
\frac{1}{2}\!\sum_{i=1,2}\!\sum_{\,\,\lambda's'r'}\!\!{\cal M}^*_{i,qg}(l)
{\cal M}_{i,qg}(\tilde{l})\! &=&\!\frac{4p_\perp^2(l_\perp\tilde{l}_\perp)}
{\alpha(\!1\!-\!\!\alpha)N^4\alpha''}\!\! \, \times  \\
& & \!\!\left[\alpha\!+\!\frac{1}{\alpha}
\!\left(\!\frac{\alpha(\!1\!\!-\!\!\alpha)N^2}{p_\perp^2}\right)^{\!\!2}
\!\!\!+\!\frac{(\!1\!\!-\!\!\alpha)^2}{\alpha}\!\left(\!1\!-\!
\frac{\alpha(\!1\!\!-\!\!\alpha)N^2}{p_\perp^2}\right)^{\!\!2}\right]
\!. \nonumber
\eea
Together with the formulae of Sections \ref{ampl} and \ref{cols} the 
remaining results of Sections \ref{lcs} and \ref{tcs} follow in a 
straightforward manner.

\end{document}